# Extending the n = 2 Ruddlesden–Popper solid solution $La_{2-2x}Sr_{1+2x}Mn_2O_7$ beyond x = 0.5: synthesis of $Mn^{4+}$–rich compounds

Julie E. Millburn, John F. Mitchell,* Dimitri N. Argyriou

*Materials Science Division, Argonne National Laboratory, Argonne, Illinois 60439 U. S. A.*

We report the synthesis and room temperature crystal structures of the heretofore unknown, metastable manganites $La_{2-2x}Sr_{1+2x}Mn_2O_{7+\delta}$ ($0.5 \leq x \leq 0.9$) via high temperature (T = 1650 °C) quenching followed by low temperature (T = 400 °C) annealing to fill oxygen vacancies; this approach enables access to the electronic, magnetic, and structural properties of previously unexplored compositions in this important CMR system.

(submitted to Chemical Communications 4/6/99)

Mixed-valent manganite perovskites have received considerable attention in recent years, primarily because of the observation of colossal magnetoresistance (CMR) and more generally due to the unusually strong coupling between their lattice, spin, and charge degrees of freedom.[1-3] Although the focus of interest has primarily rested with the pseudocubic $Ln_{1-x}A_xMnO_3$ perovskites (A = Ca, Sr, Ba, Pb, Bi), naturally layered CMR materials have also attracted attention due to their potential as model systems for low-dimensional physics[4-6] and because they exhibit substantially better low-field magnetoresistance than their three–dimensional perovskite counterparts.[7] Single phase synthesis has only been successfully achieved in the n = 2 Ruddlesden–Popper (RP) $La_{2-2x}Sr_{1+2x}Mn_2O_{7+\delta}$ manganite series for doping concentrations in the range $0.3 \leq x \leq 0.5$. Nonetheless, the structural, electronic, and magnetic phase diagram across this relatively narrow doping region is remarkably rich, displaying ferromagnetism, antiferromagnetism, canted states, spin rotations, and charge ordering.[7-11] Unfortunately, extension of the solid solution towards higher $Mn^{4+}$ concentrations has heretofore proved unsuccessful,[12] with the notable exception of the synthesis of the x = 1.0 end member compound, $Sr_3Mn_2O_{7+\delta}$.[13, 14]

Here we report on the synthesis and room temperature structural data of highly oxidized $La_{2-2x}Sr_{1+2x}Mn_2O_{7+\delta}$ phases with $0.5 \leq x \leq 0.9$. The importance of these materials for the continuing investigation and understanding of the physics of layered manganites is three–fold. First, they provide prototypes for appraisal of the interplay between aliovalent substitution and vacancy formation. Secondly, they provide models for further assessing the impact of Jahn–Teller distortions in these materials. Finally, they provide experimental data necessary to test theoretical predictions concerning the evolution of the magnetic ground state with dopant concentration.

Polycrystalline samples of $La_{2-2x}Sr_{1+2x}Mn_2O_{7+\delta}$, $0.5 \leq x \leq 0.9$, were prepared by high temperature, solid state reaction of $La_2O_3$ (Johnson–Matthey REacton 99.999%, pre–fired at 1000 °C in air for 12 h), $SrCO_3$ (Johnson–Matthey Puratronic 99.994%, dried at 150 °C in air for 12 h) and $MnO_2$ (Johnson–Matthey Puratronic 99.999%, pre–fired at 425 °C in flowing oxygen for 6 h, then slow cooled at 1 °C min$^{-1}$ to room temperature). Stoichiometric quantities of the starting materials were mixed and fired in air as powders, at 900 °C for 24 h and then 1050 °C for a further

24 h. Samples were then pressed into 13 mm diameter pellets at 6, 000 lbs. and ramped at 5 °C min$^{-1}$ to 1650 °C. After 18 h each compound was quenched directly from the synthesis temperature into dry ice. As is the case for $Sr_3Mn_2O_{7+\delta}$[2, 14], the materials are metastable below 1650 °C and must be rapidly cooled to below 1000 °C to prevent decomposition into non–RP phases. The black, poorly sintered, products were subsequently annealed for 12 h at 400 °C in flowing oxygen. Materials were reground between each firing. The $Mn^{4+}$ content of the as–made and annealed materials was determined by iodometric titration against a standardized potassium thiosulfate solution. For each composition titrations were repeated several times to ensure accurate and consistent results (typically less than a 2% spread). Approximately 50 mg portions of as–made material were rapidly heated on a thermogravimetric analysis (TGA) balance to 400 °C in flowing oxygen, held at this temperature until no further weight change was observed, then quickly cooled to room temperature. Room temperature time–of–flight (TOF) powder neutron diffraction data were collected for the oxygen annealed samples on the Special Environment Powder Diffractometer (SEPD) at Argonne National Laboratory's Intense Pulsed Neutron Source (IPNS). All crystal structure analysis was performed by the Rietveld method using the GSAS program suite.[15]

The results of the titrations and the room temperature $c/a$ ratio for the annealed samples as a function of $x$ are plotted in Fig. 1. The individual lattice parameters for selected compositions are listed in Table 1. Data included for the x = 1.0 composition, $Sr_3Mn_2O_{7+\delta}$, are taken from Ref.[13] Within the resolution of the SEPD instrument (±145 ° 2θ Δd/d (FWHM) = 0.0034) we find no evidence in our samples of the subtle phase separation previously reported for the x = 0.5 composition.[4]

As shown in Table 1, the $a$ lattice parameter monotonically decreases by 1.94% between $0.5 \leq x \leq 1.0$. In contrast, the $c$ –axis shows a minimum at $x \approx 0.65$. This results in a smooth rise in the $c/a$ ratio with increasing $x$, see Fig. 1, a complete reversal of what is observed with increasing $Mn^{4+}$ concentration in the $0.3 \leq x \leq 0.5$ region. This transformation in behavior may reflect changes in the nature of orbital occupation and/or A–type ordering in this system as a function of manganese oxidation state. Theoretical predictions of the magnetic ground state by

Maezono et al.,[16] for instance, depend upon a decreasing $c/a$ ratio. In light of the present results, such calculations may require revision.

The as-made materials are found by titration to display an average Mn oxidation state of approximately +3.5 irrespective of composition. Analogy with $Sr_3Mn_2O_{7+\delta}$[13] suggests this to be a consequence of the formation of oxygen vacancies in the materials during synthesis. Thermogravimetric analysis corroborates this hypothesis; illustrative results for the x = 0.75 composition are shown in Fig. 2. The observed weight gain indicates that a substantial quantity of oxygen is incorporated by the sample upon heating. Assuming the end–point corresponds to the stoichiometric material, $La_{0.5}Sr_{2.5}Mn_2O_7$, then the as–made compound may be formulated as $La_{0.5}Sr_{2.5}Mn_2O_{6.5}$ i. e. $\delta$ = -0.24. Since, the average manganese oxidation is given by $3 + x + \delta$, this value of $\delta$ corresponds to an oxidation state of +3.51(2) in good agreement with the titrimetric result of +3.48(2). Fig. 1 shows the monotonic decrease in $\delta$ (or increase in number of oxygen vacancies formed during synthesis) with $x$ that gives rise to the constant, dopant independent Mn oxidation state. Once annealed at 400 °C in oxygen, all the materials are found to be stoichiometric, $\delta$ = 0.00(2). This is corroborated by refinement of the oxygen content in the powder neutron diffraction data: all oxygen site occupancies are unity within error. These findings suggest that the initial synthesis conditions impose an upper limit on the transition–metal oxidation state. In fact, in our experience, $Mn^{+3.5}$ is the maximum mean oxidation state attainable in a layered n = 2 manganite system by any high temperature route. Prior attempts to extend the solid solution beyond $x$ = 0.5 in this and other $Ln_{2-2x}Sr_{1+2x}Mn_2O_7$ series by conventional high temperature solid state techniques results in the formation of mixed–phase materials.[12]

The use of a two–step procedure such as reported here, in which formation of vacancies within the structure is followed by low temperature oxygen annealing to yield the stoichiometric material, thus seems essential in providing access to compositions containing higher concentrations of $Mn^{4+}$. Access to these compositions will permit a more complete investigation of the interplay among structure, magnetism, and transport in layered manganites. Full characterization of the

temperature–dependent crystallographic and magnetic structure of the compositions prepared here will be reported in a subsequent paper.

The authors thank Simine Short for assistance with the TOF powder neutron diffraction measurements. This work was supported by the U. S. Department of Energy, Basic Energy Sciences–Materials Sciences under contract W–31–109–ENG–38 (JEM, JFM) and W–7405–ENG–36 (DNA).

| $x$ | $a$ / Å | $c$ / Å |
|---|---|---|
| 0.50 | 3.87470(3)[a] | 20.0031(3) |
| 0.55 | 3.87035(4) | 19.9862(3) |
| 0.60 | 3.86490(3) | 19.9784(3) |
| 0.75 | 3.84317(5) | 19.9968(3) |
| 0.80 | 3.83533(5) | 20.0070(4) |
| 0.90 | 3.81834(5) | 20.0325(5) |
| 1.00 | 3.79972(5) | 20.0959(4) |

Table 1: Room temperature lattice parameters and unit cell volume, from powder neutron diffraction data, for $La_{2-2x}Sr_{1+2x}Mn_2O_7$, $0.5 \leq x \leq 0.9$. [a]Numbers in parentheses are e.s.d.s.

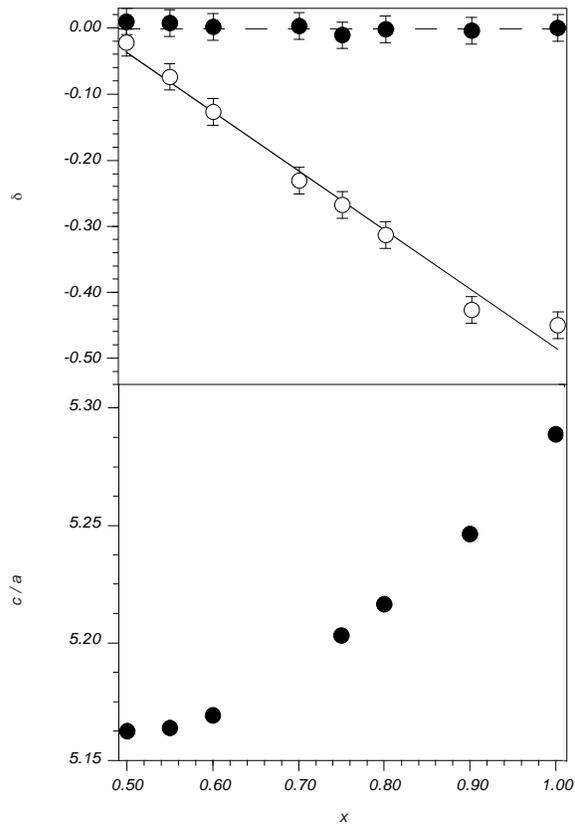

Figure 1: Oxygen non-stoichiometry, $\delta$, and $c/a$ versus $x$ in $La_{2-2x}Sr_{1+2x}Mn_2O_{7+\delta}$, $0.5 \leq x \leq 0.9$. Open spheres represent data for the as–made, 1650 °C quenched samples, closed circles that for the low temperature oxygen annealed materials.

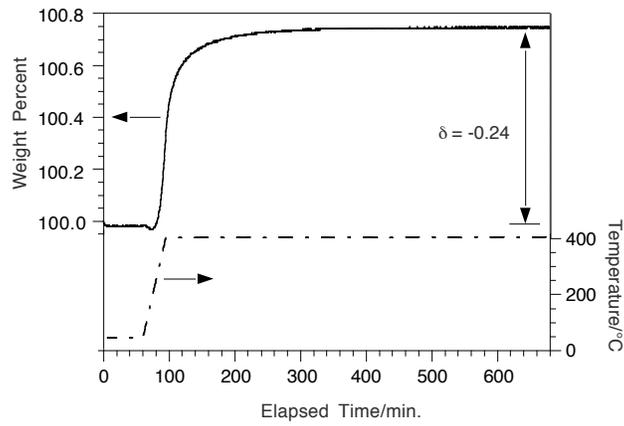

Figure 2: Thermogravimetric analysis of as–made La$_{0.5}$Sr$_{2.5}$Mn$_2$O$_{7+\delta}$ under a flowing oxygen atmosphere.